\documentclass[aps,prl,twocolumn,nofootinbib,longbibliography]{revtex4-2}

\usepackage{amsmath,amssymb}
\usepackage{mathtools}
\usepackage[dvipsnames]{xcolor}
\usepackage{hyperref}
\usepackage{microtype}
\hypersetup{
  colorlinks=true,
  linkcolor=black,
  citecolor=black,
  urlcolor=black
}
\usepackage{bm}
\usepackage{mathrsfs}
\usepackage{tikz}
\usetikzlibrary{shapes.geometric, arrows, positioning, calc, decorations.pathmorphing, shadings}
\allowdisplaybreaks[1]

\newcommand{\SU}{\mathrm{SU}}

\newcommand{\cH}{\mathcal{H}}
\newcommand{\cV}{\mathcal{V}}
\newcommand{\DD}{\mathrm{D}}
\newcommand{\ii}{\mathrm{i}}

\newcommand{\Id}{\mathbf{1}}
\newcommand{\Tr}{\mathrm{Tr}}
\newcommand{\Dom}{\operatorname{Dom}}

\newcommand{\spec}{\operatorname{spec}}

\newcommand{\norm}[1]{\left\lVert #1 \right\rVert}
\newcommand{\opnorm}[1]{\left\lVert #1 \right\rVert_{\mathrm{op}}}

\begin{document}

\title{Spectral Admissibility of Real Observers in Euclidean de Sitter Gravity}

\author{Ricardo Esp\'indola$^{1}$, \quad Ahmed Farag Ali$^{2,3}$}
\affiliation{$^{1}$Institute for Advanced Study, Tsinghua University, Beijing 100084, China}
\affiliation{$^{2}$Essex County College, Newark, NJ 07102, USA}
\affiliation{$^{3}$Department of Physics, Benha University, 13511 Benha, Egypt}


\begin{abstract}
The Euclidean de Sitter path integral contains the familiar phase associated with conformal negative modes. Maldacena's construction shows that a suitably included real observer can reorganize the refined state-counting problem. This paper does not rederive that cancellation. It addresses the prior semiclassical admissibility question: which observer sectors couple to the de Sitter saddle as genuine metric observers without becoming spectators or producing infrared-singular backreaction? On $S^D$, after gauge fixing and zero-mode projection, the observer's quadratic influence is governed by a Schur complement. We formulate a form-domain criterion: if the observer kernel is positive and the mixed metric-observer source is bounded after applying $\Delta_{\Phi\Phi}^{-1/2}$, the induced metric correction is a bounded quadratic-form perturbation on the chosen channel. In the gapped case, $\Delta_{\Phi\Phi}\geq m_*^2\mathbf{1}$ gives $\|K^\dagger \Delta_{\Phi\Phi}^{-1} K\|_{\rm op} \leq \|K\|_{\rm op}^2/m_*^2$; metric-coupled soft modes produce corrections growing as $1/\varepsilon$. We prove a sufficiency theorem: on any stable channel with coercive form $Q_{gg}^P \geq \delta_P \|h\|^2$, the Gaussian saddle remains controlled whenever $\|\Delta_{\Phi\Phi}^{-1/2} \mathfrak{j}_P\|_{\rm op}^2 < \delta_P$. We construct a localized gapped clock-detector with internal oscillators on a smeared worldline that satisfies the criterion with a computable bound and gives explicit $S^4$ benchmark versus the round-sphere TT scale. The conformal channel is treated only as an indefinite or contour-defined sector; boundedness does not imply positivity. The criterion identifies the semiclassically admissible observer class. Phase cancellation follows only when this class overlaps the relevant conformal or negative-mode sector and is combined with an independent contour or state-counting prescription.
\end{abstract}

\maketitle

\paragraph{Introduction.---}
The modern Euclidean de Sitter problem begins with a tension between thermodynamics and the gravitational path integral. Gibbons and Hawking showed that de Sitter space has a cosmological horizon with a temperature and entropy, placing it within gravitational thermodynamics \cite{GibbonsHawking1977}. Gibbons, Hawking, and Perry then emphasized that the Euclidean Einstein--Hilbert action is not positive definite: the conformal factor enters with the wrong sign, so the functional integral is structurally indefinite \cite{GibbonsHawkingPerry1978}. Polchinski sharpened this issue by computing the phase of the sphere saddle, obtaining in his conventions a factor \((-\ii)^{d+2}\) in the sum over spheres \cite{Polchinski1989}. In current de Sitter discussions the same obstruction is often written as \(\ii^{D+2}\), depending on contour and sign conventions. In either language, the central point is the same: the Euclidean sphere partition function is not automatically a manifestly positive state-counting object.

Further clarification came from systematic analyses of sphere path integrals and quantum de Sitter entropy. Law developed a covariant treatment of local one-loop path integrals on spheres for fields of arbitrary spin, including higher-spin analogs of Polchinski's phase \cite{Law2021SphereInts}. Anninos, Denef, Law, and Sun computed exact one-loop corrected de Sitter entropy and related the sphere partition function to a quotient of quasicanonical bulk and Euclidean edge partition functions \cite{AnninosEtAl2022QdS}. These works clarified the technical meaning of the Euclidean sphere computation, but they did not by themselves remove the interpretational tension created by its phase.

A qualitatively new step was taken by Maldacena, who argued that once one includes a suitably localized observer, the troublesome phase is mostly cancelled in the state-counting quantity, although an overall minus sign remains \cite{Maldacena2024RealObservers}. This result is the starting point rather than the target of the present paper. Maldacena shows that a suitable real observer can change the relevant state-counting phase; the question addressed here is which observer sectors are semiclassically admissible as metric-coupled observers before such a contour or state-counting mechanism is invoked. A recent companion analysis sharpened this issue by distinguishing gravitational observers from topological spectators and by showing that an information-bearing clock is necessary but not sufficient unless its fluctuations overlap the relevant conformal negative modes \cite{Ali2026RealObserversResolve}. In this sense the present work supplies an infrared admissibility criterion for Maldacena-type observers, not an alternative derivation of the Maldacena cancellation. Closely related questions about the phase have since been developed in several directions. Ivo, Maldacena, and Sun studied phases on spaces \(S^p\times M_q\) and connected part of the additional phase to negative-mass-squared modes visible after Kaluza--Klein reduction \cite{IvoMaldacenaSun2025}. Shi and Turiaci formulated a method for computing phases of gravitational path integrals on Einstein spaces and found, for example, that \(S^2\times S^{D-2}\) is real and positive \cite{ShiTuriaci2025PhaseGPI}. Chen, Stanford, Tang, and Yang revisited the phase of the de Sitter density of states for a charged black-hole observer in a reduced model and argued for a positive final density of states after reconsidering the state-counting prescription \cite{ChenEtAl2025PhaseDensity}.

The observer question has also been sharpened in algebraic and path-integral frameworks \cite{ChandrasekaranLongoPeningtonWitten2023,AbdallaAntoniniIliesiuLevine2025,HarlowUsatyukZhao2025,Chen2025ObserversHilbert,AkersEtAl2025HolographicMaps,ChenXu2025CovariantObservers,BlommaertKudlerFlamUrbach2025,HorowitzMarolfSantos2025Constraints,AguilarGutierrezBahiruEspindola2024CentaurAlgebra,EspindolaMiyashita2025FlowMicrostates}. In particular, the centaur-algebra construction shows how flow geometries interpolating between asymptotic AdS and interior de Sitter regions can drive the algebra of observables toward a type-II description through finite-cutoff constraints and infalling-observer protocols, while the flow-geometry microstate construction gives a holographically controlled realization of de Sitter microstates through particles inserted behind the de Sitter horizon and wormhole contributions. Taken together, these works emphasize observer algebras, observer-relative Hilbert spaces, de Sitter microstates, and observer-centered state preparation. The present paper is complementary to that literature. It does not attempt to derive the full nonperturbative observer algebra or reproduce the contour analysis of a specific observer construction. It isolates a prior semiclassical question about the quadratic response of the de Sitter saddle itself: after one knows that an appropriate observer can reorganize the phase, what spectral conditions distinguish an appropriate gravitational observer from an information-bearing spectator or an infrared-singular sector?

The distinction at stake is simple. A sector may carry an internal Hilbert space, store records, or function as a clock, while still remaining gravitationally irrelevant in the infrared. Conversely, a comparatively simple sector can become gravitationally important if its soft modes couple to the metric. The decisive quantity is therefore not information in the abstract, but the observer's \emph{metric-coupled infrared susceptibility}. In a Gaussian expansion this susceptibility enters through the inverse observer kernel and hence through a Schur complement.

The contribution of this paper is to formulate that criterion in a way that is mathematically precise enough to expose the necessary assumptions. The main estimate is elementary as an operator inequality; its use here is not to introduce a new theorem of functional analysis, but to identify the exact spectral quantity controlling observer relevance in the de Sitter saddle. We make this explicit at the level of closed quadratic forms, including the possibility that the mixed coupling is naturally a form-bounded source rather than a bounded \(L^2\to L^2\) operator. This refinement is important for matter models whose metric variation contains derivatives of the observer fluctuation. We then show how the criterion reduces to the familiar \(\opnorm{K}^2/m_*^2\) bound in the operator-bounded case, how it fails for metric-coupled soft modes, and how it appears in a solvable scalar response model on \(S^D\). Finally, we go beyond a purely necessary condition in two limited but concrete ways: we prove a sufficiency theorem for Gaussian control on stable positive channels, and we construct a localized gapped clock-detector model whose metric source satisfies the theorem with an explicit bound, including a concrete \(S^4\) benchmark and a determinant-sector check. The criterion is therefore a spectral eligibility test for controlled observer coupling. It supplies the infrared input on which the separate conformal-contour or state-counting part of Maldacena-type phase arguments acts.

\paragraph{Informational sectors versus gravitational relevance.---}
Relational and algebraic approaches to time have long emphasized that a clock is a subsystem whose internal states support an intrinsic ordering of events \cite{PageWootters1983,ConnesRovelli1994}. Many physical systems realize forms of stable ordering or memory. Topologically ordered phases support protected logical sectors through braiding and ground-state degeneracy \cite{WenNiu1990QHTopDegeneracy,Kitaev2003ToricCode,QiHughesZhang2008TIFieldTheory}. Defect networks in cosmology and condensed matter carry robust geometric histories through topology and reconnection processes \cite{Kibble1976TopologyStrings,VilenkinShellard1994,Zurek1985CosmoSuperfluidHelium}. These examples show that information storage and logical ordering are abundant in physics. They also show that information, by itself, does not determine whether a sector is gravitationally relevant to the de Sitter phase problem.

The missing quantity is infrared response in the metric channel. A sector with a rich internal state space may be an exact or perturbative spectator if its mixed quadratic block with the metric vanishes or remains small. A sector with fewer internal states may be dynamically dangerous if its low-energy modes overlap a gravitational channel. Observation becomes gravitationally meaningful only when the sector has finite but non-negligible susceptibility to the metric fluctuations that control the saddle.

\paragraph{Projected quadratic forms and channel convention.---}
Consider the gauge-fixed Euclidean path integral for gravity coupled to an observer sector \(\Phi\) on the de Sitter sphere \(S^D\),
\begin{equation}
  Z_{\mathrm{tot}}
  =
  \int \DD g\,\DD\Phi\;
  \exp\!\big[-I_{\mathrm{grav}}[g]-I_{\mathrm{obs}}[g,\Phi]\big].
\end{equation}
We expand around a background configuration \((\bar g,\bar\Phi)\) satisfying the renormalized saddle equations,
\[
g=\bar g+h,
\qquad
\Phi=\bar\Phi+\varphi .
\]
Purely local observer contributions, including vacuum-energy shifts and curvature counterterms, are absorbed into the renormalized background and into the local part of the gravitational Hessian. The Schur term studied below is therefore the \emph{mixed-block} contribution to the Gaussian influence functional. In a complete one-loop treatment there are also determinant contributions such as \(\frac12\Tr\log\Delta_{\Phi\Phi}[g]\). Those terms must be renormalized or analyzed separately. When the observer kernel develops near-zero eigenvalues, variations of the determinant contain the same inverse-kernel structure schematically through terms of the form \(\Tr(A^{-1}\delta A)\) and \(\Tr(A^{-1}\delta A A^{-1}\delta A)\). Thus determinant terms can also inherit infrared enhancement from soft metric-coupled modes. The present paper isolates the mixed-block Schur contribution, but the same susceptibility logic gives the necessary check for the determinant sector: no part of the one-loop influence functional should contain an unbounded response on the metric channel under consideration.

After gauge fixing and projection away from exact gauge modes and exact global zero modes, let \(P\cH_g\) be a chosen projected metric channel. The projector \(P\) may represent a transverse-traceless channel, a scalar channel, or the conformal channel after the relevant contour or indefinite-form convention has been specified. This channel language is essential. On stable channels the gravitational quadratic form is represented by a closed semibounded form. On the conformal channel the Euclidean gravitational form is not positive; throughout this paper the Schur estimates in that channel mean only boundedness of the observer-induced perturbation of the gauge-fixed indefinite or contour-defined Hessian. They do not imply positivity and do not by themselves solve the conformal-factor problem.

Let \(\cH_\Phi\) be the Hilbert space for observer fluctuations and let
\begin{equation}
  A\equiv \Delta_{\Phi\Phi}
\end{equation}
be the positive self-adjoint operator associated with the observer quadratic form on the projected observer space. We write
\begin{equation}
  \cV_\Phi=\Dom(A^{1/2}),\qquad
  \norm{\varphi}_{A}^{2}:=\norm{A^{1/2}\varphi}_{\Phi}^{2}.
\end{equation}
For a strictly positive \(A\), this is equivalent to the standard graph norm. We use the canonical Gelfand triple \(\cV_\Phi\subset\cH_\Phi\subset\cV_\Phi^*\), so that \(A^{-1/2}\) is understood as the continuous map from the appropriate subspace of \(\cV_\Phi^*\) into \(\cH_\Phi\). The mixed quadratic term is treated as a form
\begin{equation}
  B_P:P\cV_g\times \cV_\Phi\longrightarrow \mathbb C,
\end{equation}
where \(P\cV_g\) is the form domain of the chosen metric channel. For fixed \(h\in P\cV_g\), the mixed form defines a source
\begin{equation}
  \mathfrak j_P h\in \cV_\Phi^*,
  \qquad
  B_P(h,\varphi)=\langle \mathfrak j_P h,\varphi\rangle_{\cV_\Phi^*,\cV_\Phi}.
  \label{eq:j-def}
\end{equation}
This formulation is slightly more general than the common block-matrix notation because it allows the metric variation of the observer action to contain derivatives of \(\varphi\). For notational simplicity we write the real-field form; in the complex case, \(B_P\) should be replaced by the Hermitian real part of the corresponding sesquilinear pairing. In the special case where the source is represented by an \(L^2\)-bounded operator \(K_P:P\cH_g\to \cH_\Phi\), Eq.~\eqref{eq:j-def} reduces to
\begin{equation}
  B_P(h,\varphi)=\langle K_Ph,\varphi\rangle_\Phi .
\end{equation}

The quadratic action on the chosen channel is then
\begin{equation}
Q_P[h,\varphi]
=
\frac12 Q_{gg}^{P}[h]+B_P(h,\varphi)+\frac12\norm{A^{1/2}\varphi}_{\Phi}^{2}.
\label{eq:qform-block-revised}
\end{equation}
Completing the square in the observer variable gives
\begin{equation}
Q_{\rm eff}^{P}[h]
=
\frac12 Q_{gg}^{P}[h]
-
\frac12\norm{A^{-1/2}\mathfrak j_P h}_{\Phi}^{2},
\label{eq:qeff-form}
\end{equation}
whenever \(A^{-1/2}\mathfrak j_P h\) is well defined in \(\cH_\Phi\). This is the form-domain version of the formal Schur complement
\begin{equation}
  \Delta_{gg}^{\mathrm{eff},P}
  =
  \Delta_{gg}^{P}
  -
  K_P^{\dagger}A^{-1}K_P .
  \label{eq:schur}
\end{equation}
Thus the observer modifies the gravitational saddle through two ingredients: the size and channel overlap of the mixed source, and the infrared behavior of \(A^{-1}\).

\paragraph{Bounded susceptibility criterion.---}
The precise control statement is the following.

\emph{Proposition 1 (form-domain Schur bound).---}
Let \(A=\Delta_{\Phi\Phi}\) be strictly positive and self-adjoint on \(\cH_\Phi\), with form domain \(\cV_\Phi=\Dom(A^{1/2})\). Let \(P\) be a projected metric channel and suppose that the mixed source \(\mathfrak j_P:P\cV_g\to \cV_\Phi^*\) satisfies
\begin{equation}
  T_P:=A^{-1/2}\mathfrak j_P
\end{equation}
and that \(T_P\) extends to a bounded operator
\begin{equation}
  T_P:P\cH_g\longrightarrow \cH_\Phi .
\end{equation}
Then the observer-induced correction
\begin{equation}
  \mathcal S_P[h]:=\norm{A^{-1/2}\mathfrak j_P h}_{\Phi}^{2}
\end{equation}
is a bounded nonnegative quadratic form on \(P\cH_g\), with
\begin{equation}
  0\le \mathcal S_P[h]\le \opnorm{T_P}^{2}\norm{Ph}_{g}^{2}.
  \label{eq:form-schur-bound}
\end{equation}
Consequently, the effective metric quadratic form differs from the bare channel form by a bounded form perturbation:
\begin{equation}
  Q_{\rm eff}^{P}[h]
  =
  \frac12 Q_{gg}^{P}[h]
  -
  \frac12 \mathcal S_P[h].
\end{equation}

\emph{Proof.---}
By hypothesis \(T_P=A^{-1/2}\mathfrak j_P\) is bounded from \(P\cH_g\) to \(\cH_\Phi\). Hence
\[
\mathcal S_P[h]=\norm{T_P h}_{\Phi}^{2}
\le \opnorm{T_P}^{2}\norm{Ph}_{g}^{2}.
\]
The stated form perturbation follows immediately from completing the square in Eq.~\eqref{eq:qform-block-revised}. \hfill \(\square\)

The familiar operator-bound form is recovered as a corollary.

\emph{Corollary 1 (gapped operator-bounded case).---}
Assume that \(\mathfrak j_P h=K_Ph\) with \(K_P:P\cH_g\to\cH_\Phi\) bounded, and suppose
\begin{equation}
  A=\Delta_{\Phi\Phi}\ge m_*^2\Id,
  \qquad m_*>0 .
  \label{eq:gap}
\end{equation}
Then
\begin{equation}
\begin{aligned}
\opnorm{A^{-1}}
&\le m_*^{-2},
\\
\opnorm{A^{-1/2}K_P}
&\le \frac{\opnorm{K_P}}{m_*},
\\
\opnorm{K_P^\dagger A^{-1}K_P}
&\le \frac{\opnorm{K_P}^{2}}{m_*^{2}},
\end{aligned}
\label{eq:bound-op-revised}
\end{equation}
and therefore
\begin{equation}
\begin{aligned}
&\big|\langle Ph,K_P^\dagger A^{-1}K_PPh\rangle_g\big|
\\
&\qquad\le
\frac{\opnorm{K_P}^{2}}{m_*^2}\norm{Ph}_g^2 .
\end{aligned}
\label{eq:channel-bound}
\end{equation}

\emph{Proof.---}
The spectral theorem gives \(\spec(A)\subset[m_*^2,\infty)\), hence \(\opnorm{A^{-1}}\le m_*^{-2}\) and \(\opnorm{A^{-1/2}}\le m_*^{-1}\). The remaining estimates follow from submultiplicativity of the operator norm. \hfill \(\square\)

This estimate is not intended as a deep abstract result. Its purpose is to state, in the language of the de Sitter saddle, which quantity has to be finite before an observer sector can participate in a controlled semiclassical reorganization. The quantity is not the number of internal observer states, nor the existence of a clock variable, but the boundedness of \(A^{-1/2}\mathfrak j_P\): the metric-coupled infrared susceptibility of the observer on the chosen gravitational channel.

For stable projected channels, Eq.~\eqref{eq:form-schur-bound} gives an ordinary bounded perturbation estimate. For the conformal channel, the same equation remains only a boundedness statement for the induced term relative to the selected indefinite or contour-defined quadratic structure. Since the Schur correction enters with a fixed sign in the Euclidean Gaussian completion, a finite correction cannot by itself convert the conformal-factor problem into a positivity theorem. This is why the present result should be read as a necessary control criterion, not as a replacement for Maldacena-type contour or state-counting arguments \cite{Maldacena2024RealObservers}.

The previous proposition is a necessary boundedness criterion. On stable channels one can also state a limited sufficiency theorem for Gaussian control.

\emph{Proposition 2 (sufficiency for stable-channel Gaussian control).---}
Let $P\cH_g$ be a projected stable metric channel. Assume that the bare gravitational channel form is closed and coercive,
\begin{equation}
  Q_{gg}^{P}[h]
  \ge
  \delta_P\norm{Ph}_{g}^{2},
  \qquad
  \delta_P>0,
  \label{eq:bare-coercive}
\end{equation}
and that the observer source satisfies Proposition~1 with bounded susceptibility $T_P=A^{-1/2}\mathfrak j_P$. If
\begin{equation}
  \opnorm{T_P}^{2}<\delta_P,
  \label{eq:suff-smallness}
\end{equation}
then the effective metric form is closed and coercive on the same channel, with
\begin{equation}
  Q_{\rm eff}^{P}[h]
  \ge
  \frac12\big(\delta_P-\opnorm{T_P}^{2}\big)
  \norm{Ph}_{g}^{2} .
  \label{eq:suff-bound}
\end{equation}
Consequently the Gaussian expansion on that stable channel remains uniformly controlled. In the operator-bounded gapped case this sufficient condition becomes
\begin{equation}
  \frac{\opnorm{K_P}^{2}}{m_*^2}<\delta_P .
  \label{eq:suff-gap}
\end{equation}

\emph{Proof.---}
Using Eq.~\eqref{eq:qeff-form}, Proposition~1, and Eq.~\eqref{eq:bare-coercive},
\[
Q_{\rm eff}^{P}[h]
=\frac12Q_{gg}^{P}[h]-\frac12\mathcal S_P[h]
\ge
\frac12(\delta_P-\opnorm{T_P}^{2})\norm{Ph}_{g}^{2}.
\]
The perturbation is bounded relative to the Hilbert norm, so adding it to the closed form $Q_{gg}^{P}$ preserves closedness; the displayed inequality gives coercivity. \hfill $\square$

This proposition is deliberately weaker than a solution of the de Sitter phase problem. It says that, on a stable positive channel, bounded susceptibility together with the strict inequality \eqref{eq:suff-smallness} is sufficient for Gaussian control. It does not apply directly to the unrotated conformal direction, where $Q_{gg}^{P}$ is not coercive. If a contour prescription or state-counting construction replaces the conformal direction by a positive comparison form, then the same proposition can be applied to that contour-defined channel, but the contour input is external to the theorem.

\paragraph{Soft modes and loss of uniform control.---}
The opposite regime is equally important. Consider a family of observer kernels \(A_\alpha\) and mixed blocks \(K_{\alpha,P}\) in the operator-bounded setting. Suppose there exists a soft spectral window \((0,\varepsilon_\alpha]\) with \(\varepsilon_\alpha\to0\), and let
\[
P_{\mathrm{soft}}^{(\alpha)}
=
\chi_{(0,\varepsilon_\alpha]}(A_\alpha)
\]
be the corresponding spectral projector. If there are normalized metric fluctuations \(u_\alpha\in P\cH_g\) such that
\begin{equation}
\begin{aligned}
\norm{P_{\mathrm{soft}}^{(\alpha)}K_{\alpha,P}u_\alpha}_{\Phi}
&\ge c>0,
\\[-1mm]
&\hspace{1.5em}\text{uniformly in }\alpha .
\end{aligned}
\label{eq:soft-overlap}
\end{equation}
then
\begin{equation}
\begin{aligned}
\langle u_\alpha,
K_{\alpha,P}^{\dagger}A_\alpha^{-1}K_{\alpha,P}u_\alpha\rangle_g
&=
\langle K_{\alpha,P}u_\alpha,
A_\alpha^{-1}K_{\alpha,P}u_\alpha\rangle_{\Phi}
\\
&\ge
\varepsilon_\alpha^{-1}
\norm{P_{\mathrm{soft}}^{(\alpha)}K_{\alpha,P}u_\alpha}_{\Phi}^{2}
\\
&\ge
\frac{c^2}{\varepsilon_\alpha}.
\end{aligned}
\label{eq:soft}
\end{equation}
Thus the induced correction becomes arbitrarily large along that channel as \(\varepsilon_\alpha\to0\).

This is the precise sense in which metric-coupled soft observer modes destroy uniform Gaussian control. It is not enough for the observer sector to possess soft modes; those modes must overlap the metric source. Exact spectators with \(K_{\alpha,P}=0\), or sectors whose soft modes are projected away from the metric-response channel, do not trigger the divergence in Eq.~\eqref{eq:soft}.

A de Sitter-specific subtlety is worth emphasizing. Because \(S^D\) is compact, a free massless scalar on a fixed sphere does not produce an infrared catastrophe after the constant zero mode is projected out: the first nonzero scalar Laplacian eigenvalue is \(D/L^2\). Uniform loss of control therefore arises across families: for example in a decompactification limit \(L\to\infty\), in a near-critical family with \(m_*^2\to0\), or in sectors with collective modes parametrically softer than the geometric scale \(L^{-2}\). The issue is not mere masslessness on one compact background, but parametrically enhanced metric-coupled infrared susceptibility.

For the concrete family
\begin{equation}
  A_{L,m}=-\bar\nabla^2+m^2
\end{equation}
on the scalar sector with the constant mode removed, the exact inverse norm is
\begin{equation}
  \left\lVert A_{L,m}^{-1}\right\rVert_{\perp0,\mathrm{op}}
  =
  \frac{1}{D/L^2+m^2}
  =
  \frac{L^2}{D}\,\frac{1}{1+m^2L^2/D} .
  \label{eq:soft-scaling}
\end{equation}
This formula gives the sharp scaling boundary. If \(m^2L^2/D\gg1\), the susceptibility is mass controlled, \(\opnorm{A_{L,m}^{-1}}\sim m^{-2}\). If \(m^2L^2/D\ll1\), compactness controls the fixed-sphere infrared behavior, but the bound still grows as \(L^2/D\) in a decompactification limit. More generally, for \(m_L^2\sim L^{-p}\), the inverse grows as \(L^p\) for \(p<2\) and saturates at the geometric growth \(L^2/D\) for \(p\ge2\). Thus the criterion is quantitative: uniform control is lost precisely when the metric-coupled Schur coefficient inherits one of these unbounded scalings.

\paragraph{Scalar matter and the form-bound source.---}
The form-domain language is not cosmetic; it is forced by standard matter variations. Consider a minimally coupled scalar matter field \(\psi\) with quadratic action
\begin{equation}
I_\psi[g,\psi]=\frac12\int_{S^D}\!\sqrt{g}\,
\Big(g^{ab}\nabla_a\psi\nabla_b\psi+M^2\psi^2\Big).
\label{eq:scalar-matter-action}
\end{equation}
Expanding about a smooth background \((\bar g,\bar\psi)\) with \(g=\bar g+h\) and \(\psi=\bar\psi+\varphi\), the observer kernel on the matter fluctuation is
\begin{equation}
A=-\bar\nabla^2+M^2,
\label{eq:scalar-kernel-derived}
\end{equation}
and the mixed quadratic term is

\begin{multline}
B(h,\varphi)
=
\frac12\int_{S^D}\!\sqrt{\bar g}\,
h^{ab}
\Big[
-2\nabla_a\bar\psi\,\nabla_b\varphi
\\
+\bar g_{ab}
\big(
\nabla^c\bar\psi\,\nabla_c\varphi
+
M^2\bar\psi\,\varphi
\big)
\Big].
\label{eq:derived-mixed}
\end{multline}
Because \(S^D\) is compact and \(\bar\psi\) is smooth, H\"older and Sobolev estimates imply
\begin{equation}
|B(h,\varphi)|
\le
C_{\bar\psi,M}\,
\norm{h}_{L^2}
\norm{\varphi}_{H^1} .
\label{eq:derived-bounded}
\end{equation}
Here and below, unlabelled Sobolev norms in this paragraph are taken on $S^D$.
For \(A=-\bar\nabla^2+M^2\) with \(M^2>0\), the form domain is \(H^1(S^D)\). Therefore the mixed term defines
\begin{equation}
  \mathfrak j h\in H^{-1}(S^D),
  \qquad
  \norm{\mathfrak j h}_{H^{-1}}\le C_{\bar\psi,M}\norm{h}_{L^2}.
\end{equation}
Equivalently,
\begin{equation}
  \norm{A^{-1/2}\mathfrak j h}_{L^2}
  \le
  C'_{\bar\psi,M}\norm{h}_{L^2}.
  \label{eq:scalar-form-bound}
\end{equation}
Thus a standard localized matter sector naturally supplies a bounded \(A^{-1/2}\mathfrak j\), even when the mixed source is better viewed as an element of the dual form domain rather than an \(L^2\) function. This is precisely the situation covered by Proposition~1.

The same estimate also closes the loop for the trace/conformal source at the level of boundedness. If the projected metric variation is restricted to a pure trace direction,
\begin{equation}
  h_{ab}=\frac{1}{D}\,\sigma\,\bar g_{ab},
\end{equation}
then Eq.~\eqref{eq:derived-mixed} reduces to a source of the form
\begin{equation}
\begin{aligned}
B_{\rm tr}(\sigma,\varphi)
&=
\frac{1}{2D}\int_{S^D}\!\sqrt{\bar g}\,\sigma\,
\mathcal I_{\rm tr},
\\
\mathcal I_{\rm tr}
&=(D-2)\nabla^c\bar\psi\nabla_c\varphi
\\
&\quad +DM^2\bar\psi\,\varphi .
\end{aligned}
\label{eq:trace-source}
\end{equation}
For smooth \(\bar\psi\) on compact \(S^D\),
\begin{equation}
  |B_{\rm tr}(\sigma,\varphi)|
  \le C_{\bar\psi,M,D}\,
  \norm{\sigma}_{L^2(S^D)}\norm{\varphi}_{H^1(S^D)} .
  \label{eq:trace-form-bound}
\end{equation}
Thus the trace/conformal source obeys the same form-bound hypothesis as the general scalar matter source. This statement is deliberately only a boundedness result: because the gravitational conformal quadratic form is indefinite before contour choice, Eqs.~\eqref{eq:trace-source}--\eqref{eq:trace-form-bound} do not imply conformal-sector positivity. They show only that a natural scalar observer source can satisfy the required Schur-control condition on the trace channel.

\paragraph{A solvable de Sitter response model.---}
For calculational transparency we now pass to a deliberately minimal effective response model. It is not intended as a microscopic clock, detector, or full observer construction. Its role is to provide a solvable realization of the spectral mechanism. A realistic observer sector would additionally require localized internal degrees of freedom, finite response time, a physical preparation prescription, and nonzero projected overlap with the gravitational channel relevant to the phase problem.

Consider a scalar observer mode \(\phi\) on a sphere of radius \(L\), with

\begin{multline}
I^{(2)}_{\rm obs}[h,\phi]
=
\frac12\int_{S^D}\!\sqrt{\bar g}\,
\phi\,(-\bar\nabla^2+\mu^2)\phi
\\
+\gamma\int_{S^D}\!\sqrt{\bar g}\,
W^{ab}(x)h_{ab}(x)\phi(x).
\label{eq:explicit-model}
\end{multline}
where \(W^{ab}(x)\) is a fixed smooth symmetric response tensor supported in the observer region. In this simplified model the mixed source is represented by the bounded multiplication operator
\begin{equation}
  (K_Ph)(x)=\gamma\,W^{ab}(x)(Ph)_{ab}(x).
  \label{eq:Kh-explicit}
\end{equation}
Define
\begin{equation}
\norm{W}_{L^\infty,\mathrm{op}}
:=
\operatorname*{ess\,sup}_{x\in S^D}
\norm{W(x)}_{\mathrm{End}(S^2T_x^*S^D)} .
\label{eq:W-norm-def}
\end{equation}
Then
\begin{equation}
\norm{W^{ab}(Ph)_{ab}}_{L^2(S^D)}
\le
\norm{W}_{L^\infty,\mathrm{op}}
\norm{Ph}_{L^2(S^D)} .
\label{eq:W-bound}
\end{equation}

Expanding \(\phi\) in scalar spherical harmonics,
\begin{equation}
  \phi(x)=\sum_{\ell,n}\phi_{\ell n}Y_{\ell n}(x),
  \qquad
  -\bar\nabla^2Y_{\ell n}
  =
  \frac{\ell(\ell+D-1)}{L^2}Y_{\ell n},
\end{equation}
we obtain
\begin{equation}
  A Y_{\ell n}
  =
  \left[
    \frac{\ell(\ell+D-1)}{L^2}+\mu^2
  \right]Y_{\ell n}.
  \label{eq:scalar-spec}
\end{equation}
For \(\mu^2>0\),
\begin{equation}
  \opnorm{A^{-1}}
  =
  \sup_{\ell\ge0}
  \frac{1}{\ell(\ell+D-1)L^{-2}+\mu^2}
  =
  \frac{1}{\mu^2}.
  \label{eq:scalar-gap}
\end{equation}
Integrating out \(\phi\) gives
\begin{equation}
\begin{aligned}
Q_{\rm ind}^{P}[h]
&=
-\frac{\gamma^2}{2}
\sum_{\ell,n}
\Big(
\ell(\ell+D-1)L^{-2}+\mu^2
\Big)^{-1}
\\
&\qquad\qquad\times
\big|\langle W^{ab}(Ph)_{ab},Y_{\ell n}\rangle\big|^2 .
\end{aligned}
\label{eq:scalar-schur}
\end{equation}
Therefore
\begin{equation}
\begin{aligned}
|Q_{\rm ind}^{P}[h]|
&\le
\frac{\gamma^2}{2\mu^2}
\norm{W^{ab}(Ph)_{ab}}_{L^2(S^D)}^2
\\
&\le
\frac{\gamma^2\norm{W}_{L^\infty,\mathrm{op}}^2}{2\mu^2}
\norm{Ph}_{L^2(S^D)}^2 .
\end{aligned}
\label{eq:scalar-bound}
\end{equation}
This is the explicit \(S^D\) realization of Corollary~1.

If \(\mu=0\) and the exact constant mode is projected out, then
\begin{equation}
  \norm{(-\bar\nabla^2)^{-1}}_{\perp0,\mathrm{op}}
  =
  \frac{L^2}{D}.
  \label{eq:massless-sphere}
\end{equation}
Thus a massless scalar observer on a fixed compact sphere is still infrared bounded after zero-mode removal. It becomes dangerous only across families in which \(L^2/D\) is parametrically large or in which additional collective modes soften relative to \(L^{-2}\).

\paragraph{A localized gapped clock-detector observer.---}
The scalar response model above is useful because it is solvable, but it is not itself a clock. We now give a more observer-like effective model: a localized detector with internal oscillator degrees of freedom, finite response time, and a smeared worldtube coupling to the metric. This is still a Gaussian effective observer, not a microscopic model of consciousness or a derivation of Maldacena's state-counting prescription, but it is realistic enough to show that the criterion is not restricted to bulk scalar fields.

Let \(\Gamma\subset S^D\) be a smooth closed Euclidean observer trajectory of length \(\beta_\Gamma\), parametrized by \(\tau\in[0,\beta_\Gamma]\). For a great circle on the round sphere of radius \(L\), \(\beta_\Gamma=2\pi L\); more generally, for a geometrically natural closed trajectory one has \(\beta_\Gamma=c_\Gamma L\) with \(c_\Gamma=O(1)\). Thus the nonzero Fourier modes along the detector scale with the de Sitter radius, while the internal gap \(\Omega_0\) supplies an independent finite response scale. Let \(q^I(\tau)\), \(I=1,\ldots,N\), be internal pointer or clock variables with periodic Euclidean boundary conditions. The free detector kernel is
\begin{equation}
  A_{\rm clk}
  =
  -\frac{\mathrm d^2}{\mathrm d\tau^2}\Id_N+\Omega^2\Id_N+M,
  \label{eq:clock-kernel}
\end{equation}
where \(M=M^T\) is a constant positive semidefinite matrix. If \(\Omega_0^2:=\Omega^2+\lambda_{\min}(M)>0\), then
\begin{equation}
  A_{\rm clk}\ge \Omega_0^2\Id,
  \qquad
  \opnorm{A_{\rm clk}^{-1}}\le \Omega_0^{-2} .
  \label{eq:clock-gap}
\end{equation}
Thus the clock has a finite response time of order \(\Omega_0^{-1}\).

To localize the detector without introducing distributional sources, choose smooth smearing functions \(\rho_\sigma(x;X(\tau))\) supported in a tubular neighborhood of radius \(\sigma\) around the trajectory and smooth bounded tensor profiles \(E_I^{ab}(\tau,x)\). The linear metric source seen by the clock is
\begin{multline}
  (K_{\rm clk}Ph)^I(\tau)
  =
  \lambda_I
  \int_{S^D}\!\sqrt{\bar g}\,
  \rho_\sigma(x;X(\tau))
  \\
  \times
  E_I^{ab}(\tau,x)(Ph)_{ab}(x) .
  \label{eq:clock-source}
\end{multline}
The corresponding quadratic detector action is
\begin{multline}
I^{(2)}_{\rm clk}[h,q]
=
\frac12\int_0^{\beta_\Gamma}\!\mathrm d\tau\,
q^I(A_{\rm clk}q)^I
\\
+
\int_0^{\beta_\Gamma}\!\mathrm d\tau\,
q^I(\tau)(K_{\rm clk}Ph)^I(\tau) .
\label{eq:clock-action}
\end{multline}
This is the Euclidean analogue of a localized detector: the degrees of freedom are internal and one-dimensional along the observer trajectory, while the smearing radius \(\sigma\) prevents ultraviolet singularities in the metric source.

The source map is bounded. Indeed, by Cauchy--Schwarz in the spatial variable,
\begin{equation}
  |(K_{\rm clk}Ph)^I(\tau)|
  \le
  |\lambda_I|\,
  c_I(\tau)\,
  \norm{Ph}_{L^2(S^D)},
  \label{eq:clock-cs}
\end{equation}
where
\begin{equation}
  c_I(\tau)
  :=
  \norm{\rho_\sigma(\cdot;X(\tau))E_I(\tau,\cdot)}_{L^2(S^D)} .
\end{equation}
Therefore
\begin{equation}
  \norm{K_{\rm clk}Ph}_{L^2(S^1_{\beta_\Gamma};\mathbb C^N)}
  \le
  \Lambda_{\rm clk}\norm{Ph}_{L^2(S^D)},
  \label{eq:clock-source-bound}
\end{equation}
with
\begin{equation}
  \Lambda_{\rm clk}^{2}
  :=
  \sum_{I=1}^{N}|\lambda_I|^2
  \int_0^{\beta_\Gamma}\!\mathrm d\tau\,c_I(\tau)^2 .
  \label{eq:clock-lambda}
\end{equation}
Combining Eqs.~\eqref{eq:clock-gap} and \eqref{eq:clock-source-bound} gives
\begin{equation}
  \opnorm{A_{\rm clk}^{-1/2}K_{\rm clk}}
  \le
  \frac{\Lambda_{\rm clk}}{\Omega_0} .
  \label{eq:clock-schur-bound}
\end{equation}
Hence the localized clock-detector satisfies Proposition~1. On a stable channel with spectral scale \(\delta_P\), Proposition~2 gives the explicit sufficient condition
\begin{equation}
  \frac{\Lambda_{\rm clk}^{2}}{\Omega_0^2}<\delta_P .
  \label{eq:clock-sufficiency}
\end{equation}
When the ratio \(\Lambda_{\rm clk}^{2}/\Omega_0^2\) is much smaller than \(\delta_P\), the clock is a perturbative spectator. When it is comparable to \(\delta_P\) but remains finite, it is a genuine gravitational observer relative to the chosen channel. When \(\Omega_0\to0\) at fixed source strength, the response becomes soft and the model reproduces the loss of control described by Eq.~\eqref{eq:soft}. This detector therefore supplies an explicit observer-sector realization of the paper's central criterion: finite internal dynamics, localization, and finite metric-coupled infrared susceptibility are sufficient for stable-channel Gaussian control.

A simple numerical benchmark makes the condition concrete. Take \(D=4\), a single pointer \(N=1\), \(M=0\), and a great-circle trajectory, so \(\beta_\Gamma=2\pi L\). Choose a normalized smearing and tensor profile such that \(c(\tau)=1\). If the coupling is scaled as
\begin{equation}
  \lambda = \frac{\alpha\,\Omega_0}{\sqrt{\beta_\Gamma}\,L},
  \label{eq:clock-s4-coupling}
\end{equation}
then \(\Lambda_{\rm clk}^{2}=\alpha^2\Omega_0^2/L^2\), and the clock susceptibility becomes
\begin{equation}
  \frac{\Lambda_{\rm clk}^{2}}{\Omega_0^2}
  =
  \frac{\alpha^2}{L^2} .
  \label{eq:clock-s4-susceptibility}
\end{equation}
For the round-sphere TT benchmark in \(D=4\), Eq.~\eqref{eq:ttgap} gives \(\delta_{TT}=6/L^2\). Hence this explicit detector is safely perturbative for \(\alpha^2\ll6\), reaches order-one but controlled backreaction for \(\alpha^2=O(1)\) below \(6\), and violates the sufficiency bound in the normalized TT comparison-operator units when \(\alpha^2\ge6\); restoring the full gravitational Hessian normalization, including the Einstein--Hilbert prefactor, gauge-fixing convention, and Newton constant, rescales this numerical threshold. If the normalized profile has \(c(\tau)=c_0\neq1\), the same statement holds with \(\alpha\) replaced by \(\alpha c_0\). This example shows explicitly how the abstract inequality \eqref{eq:clock-sufficiency} is read in geometric units.

The determinant sector is also harmless in this effective clock model. In the simplified construction above, the kernel \(A_{\rm clk}\) is fixed on the chosen background trajectory, so \(\delta_h A_{\rm clk}=0\) and the term \(\frac12\Tr\log A_{\rm clk}\) contributes no metric Hessian. If one keeps a mild geometric dependence of the clock kernel, for example through the induced length element or redshift along \(\Gamma\), write \(\mathcal D_{\rm clk}:=\frac12\Tr\log A_{\rm clk}\). Its first two variations have the column-safe form
\begin{equation}
\begin{aligned}
\delta \mathcal D_{\rm clk}
&=
\frac12\Tr\!\left(A_{\rm clk}^{-1}\delta A_{\rm clk}\right),
\\
\delta^2 \mathcal D_{\rm clk}
&=
\frac12\Tr\!\left(A_{\rm clk}^{-1}\delta^2 A_{\rm clk}\right)
\\
&\quad -
\frac12\Tr\!\left(
A_{\rm clk}^{-1}\delta A_{\rm clk}
A_{\rm clk}^{-1}\delta A_{\rm clk}
\right) .
\end{aligned}
\label{eq:clock-det-variation}
\end{equation}
Thus any determinant-induced metric response is controlled by the same inverse-clock scale, schematically by powers of \(\Omega_0^{-2}\), provided the geometric variations \(\delta A_{\rm clk}\) and \(\delta^2A_{\rm clk}\) are bounded trace-class form perturbations on the clock domain. Softening the clock gap would enhance both the Schur term and the determinant variation; a finite \(\Omega_0\) bounds both. Within the simplified clock model and under these trace-class boundedness assumptions, the determinant sector introduces no additional infrared obstruction. More detailed metric-dependent microscopic observer kernels are analyzed by applying the same inverse-gap test to their first and second metric variations.

\paragraph{The algebraic approach for the clock-detector observer.---}
The localized gapped clock-detector satisfies the infrared admissibility criterion and supplies the Gaussian observer input for the observer-algebra viewpoint. We now state the corresponding algebraic interpretation (for a review on the subject see \cite{Witten:2018zxz}). A sector may possess a large internal Hilbert space or serve as a clock, yet it remains gravitationally inert whenever its stress-tensor response decouples from the metric fluctuations. In a local QFT setting, such a decoupled sector belongs to the spectator part of the Type III$_1$ bulk algebra; in a finite-dimensional or quantum-mechanical clock model, it appears as a spectator factor that leaves the gravitationally dressed algebra unchanged. Conversely, a sector enters the gravitational observer construction when its metric-coupled infrared susceptibility is finite, controlled, and non-negligible on the relevant metric channel. In the Gaussian framework this susceptibility is the inverse-kernel response appearing in the Schur complement (\ref{eq:schur}). In the algebraic framework it supplies the semiclassical clock input for the crossed-product dressing and trace construction.

A particularly illuminating example of the power of these algebraic structures in quantum gravity was provided by Liu and Leutheusser \cite{Leutheusser:2021frk}. In the strict large-$N$ limit of $\mathcal{N}=4$ super Yang-Mills theory (above the Hawking-Page transition), the algebra of single-trace operators restricted to the right boundary of a two-sided black hole emerges as a ${\rm Type~III}_1$ von Neumann algebra. This algebra captures the entanglement properties of the underlying state such as the entropy divergence due to entanglement fluctuations characteristic of local quantum field theory. In the bulk, geometric concepts such as the horizon become emergent properties of the algebra. Importantly, its commutant on the left boundary is likewise ${\rm Type~III}_1$.

A crucial mathematical concept in the theory of operator algebras is the crossed-product construction. By forming the crossed product of the ${\rm Type~III}_1$ algebra with its own modular automorphism group one obtains a ${\rm Type~II}_\infty$ (or, after imposing the appropriate energy/area constraint, ${\rm Type~II}_1$) factor \cite{Witten:2021unn}. The resulting algebra is equipped with a faithful normal semifinite trace that is independent of the choice of state. In the context of quantum gravity this construction has had profound consequences in recent years: it supplies a rigorous algebraic derivation of the generalized entropy of the black-hole bifurcation surface (classical extremal surface) without any appeal to Euclidean gravity or replica tricks \cite{Chandrasekaran:2022eqq}, it makes the generalized second law a direct consequence of the monotonicity of the trace under trace-preserving inclusions \cite{Chandrasekaran:2022eqq}, and it provides a background-independent description of the algebra of observables outside a horizon \cite{Witten:2023xze}. The same logic appears in flow-geometry settings: the centaur-algebra construction identifies type-II structures generated by gravitational constraints in AdS-to-dS flow geometries and by infalling-observer protocols \cite{AguilarGutierrezBahiruEspindola2024CentaurAlgebra}, while flow-geometry microstates realize de Sitter horizon entropy through controlled microstate counting and wormhole contributions \cite{EspindolaMiyashita2025FlowMicrostates}. Precisely the same crossed-product mechanism underlies the ${\rm Type~II}_1$ algebra of gravitationally dressed observables in the de Sitter static patch \cite{Chandrasekaran:2022cip}, where the observer’s proper-time flow along its worldline---realized explicitly by the gapped clock-detector model---plays the role of the modular generator.

The gapped clock-detector model supplies precisely the required modular generator. Its Euclidean quadratic action Eq.~(\ref{eq:clock-source}) together with the gapped kernel $A_{\rm clk}$ (with $\Omega_0^2 > 0$) yields, after Wick rotation, the Hamiltonian $H_{\rm clk}$ of a collection of harmonic oscillators with frequencies bounded below by $\Omega_0$. The thermal Gibbs state $\rho_{\rm clk}$ defines a GNS representation in which the modular operator $\Delta_{\rm clk}$ acts by Heisenberg evolution on the physical copy. Gravitational dressing embeds this into the full observer algebra. The algebra $\mathcal{A}_P$ is then obtained by performing the crossed product on this projected sector and imposing the physical projector $\Pi$:
\begin{equation}
\mathcal{A}_P = \Pi \Bigl( \mathcal{A} \rtimes_\alpha \mathbb{R} \Bigr) \Pi \Bigm|_{P}.
\end{equation}
The algebraic construction mirrors the Gaussian analysis. The observer backreaction (via the clock-detector) only affects the observable degrees of freedom that lie in the selected channel $P$. By working with $\mathcal{A}_P$ we are therefore describing the algebra of observables that are gravitationally dressed and consistent with the particular quadratic form $Q_P^{gg}$ (or its effective version $Q_P^{\rm eff}$). This yields the complete modular operator associated to  $\mathcal{A}_P$
\begin{equation}
\Delta = e^{-\beta_\Gamma (H_{\rm bulk} + H_{\rm clk} + x)},
\end{equation}
where $x$ is the conjugate observer energy coordinate satisfying the Heisenberg algebra $[H_{\rm obs}, x] = i$ and $\Pi = \Theta(-H_{\rm obs} - x)$. The bounded Schur correction (Eq.~(\ref{eq:qform-block-revised})) ensures that the clock contribution is compatible with the projected Gaussian expansion. The gravitational constraint, dressing map, modular flow, and observer state preparation then complete the crossed-product trace construction.

The crossed-product trace $\operatorname{Tr}$ on $\mathcal{A}$ is faithful, semifinite, and cyclic, and satisfies the KMS condition with respect to the modular flow. For a semiclassical state $\tilde{\Phi} = \Phi \otimes f(x)$ (slowly varying clock wavefunction, $\epsilon \ll \beta_{\rm dS}$) the density matrix relative to $\operatorname{Tr}$ is, via the Connes cocycle
\begin{equation}
u_{\Phi|\Psi}(t) = \Delta_{\Phi|\Psi}^{it} \Delta_{\Psi}^{-it},
\end{equation}
given in terms of the relative modular operator $\Delta_{\Phi|\Psi}$ of the bulk algebra. The cocycle identity
\begin{equation}
\Delta_{\Phi|\Psi}^{it} = u_{\Phi|\Psi}(t) \Delta_{\Psi}^{it},
\end{equation}
then implies that the entropy of this state takes the form
\begin{equation}
\begin{aligned}
S(\rho_{\tilde{\Phi}})
&= S_{\rm gen}(\infty)-S_{\rm rel}(\Phi|\Psi)+C
\\
&= S_{\rm gen}+C .
\end{aligned}
\end{equation}
The term $S_{\rm gen}(\infty)$ arises from the horizon-area shift induced by the observer energy expectation value $\langle H_{\rm obs}\rangle$ (via the linear response of the de Sitter metric), the relative entropy $-S_{\rm rel}(\Phi \lvert\Psi)$ is the bulk contribution obtained after applying the cocycle identity to the modular operators, and $C$ is the state-independent renormalization constant inherent to the Type II$_1$ trace.

At the Gaussian level, the spectator/observer distinction is diagnosed by the bounded-susceptibility condition of Proposition~1 together with non-negligible overlap with the metric channel. This is the infrared gate through which a sector enters as a gravitational observer in the crossed-product description. Proposition~2 then supplies the stronger sufficient condition for uniform Gaussian control on stable channels. The Type II$_1$ algebra of gravitationally dressed observables is obtained when this controlled observer input is combined with the full gravitational constraint and modular crossed-product construction.

When the boundedness condition of Proposition~1 fails---for example, when soft-mode overlap produces an unbounded Schur correction $\sim 1/\varepsilon$---the Gaussian observer sector becomes infrared singular. The finite-susceptibility crossed-product interpretation is then lost at the semiclassical level unless an independent infrared regulator or nonperturbative construction replaces the Gaussian saddle. The sector may still carry internal states or function as an informational clock, but it does not qualify as a controlled gravitational dressing sector in the present criterion.
\paragraph{One-mode reduction and sign.---}
The mechanism is transparent in a single projected mode,
\begin{equation}
I^{(2)}_{\mathrm{toy}}
=
\frac12
\begin{pmatrix}
h & \phi
\end{pmatrix}
\begin{pmatrix}
\lambda_g & \beta \\
\beta & m^2
\end{pmatrix}
\binom{h}{\phi}.
\label{eq:toy}
\end{equation}
Integrating out \(\phi\) gives
\begin{equation}
  I^{(2)}_{\mathrm{eff}}[h]
  =
  \frac12
  \left(
    \lambda_g-\frac{\beta^2}{m^2}
  \right)h^2 .
  \label{eq:toy-eff}
\end{equation}
If \(m^2\) stays bounded below, the correction is finite. If \(m^2\to0\) with \(\beta\neq0\), it diverges as \(\beta^2/m^2\). If \(\beta=0\), the observer is an exact spectator.

The sign interpretation is channel dependent. If \(\lambda_g>0\) represents a stable TT-like comparison mode, then \(\beta^2/m^2\sim\lambda_g\) marks order-one backreaction on that channel. If \(\lambda_g=\lambda_{\rm conf}<0\) represents a pre-existing conformal negative direction, the same formula shows that bounded backreaction alone does not remove the wrong sign. It produces a finite shift of an indefinite quadratic form, not a positivity theorem. The conformal problem therefore requires additional contour or state-counting input beyond the spectral bound.

\paragraph{Quantitative observer regimes.---}
The scalar model gives a useful operational taxonomy on any fixed projected channel. Let \(\delta_{\mathrm{st}}>0\) denote the lower spectral bound of a stable part of the gravitational Hessian on a chosen stable subspace \(\cH_{\mathrm{st}}\subset P\cH_g\). Eq.~\eqref{eq:scalar-bound} implies that, if
\begin{equation}
  \frac{\gamma^2\norm{W}_{L^\infty,\mathrm{op}}^2}{2\mu^2}
  <
  \delta_{\mathrm{st}},
  \label{eq:scalar-perturbative}
\end{equation}
then the observer is perturbative on that stable channel: it renormalizes the quadratic form but cannot compete with the channel gap.

For comparison with the round-sphere geometry, one may introduce the representative positive TT comparison operator
\begin{equation}
\mathcal O_{TT}h_{ab}^{TT}
:=
\left(-\bar\nabla^2-\frac{2}{L^2}\right)h_{ab}^{TT}
=
\lambda^{TT}h_{ab}^{TT}.
\label{eq:ott}
\end{equation}
This operator is only a benchmark scale, not an identification of the full gauge-fixed graviton Hessian in all conventions. Using the standard TT harmonic spectrum on the round sphere \cite{Higuchi1987,RubinOrdonez1984},
\begin{equation}
\begin{aligned}
-\bar\nabla^2h_{ab}^{TT(\ell)}
&=
\frac{\ell(\ell+D-1)-2}{L^2}
 h_{ab}^{TT(\ell)},
\\
&\hspace{1.5em}\ell\ge2 .
\end{aligned}
\label{eq:ttharm}
\end{equation}
one obtains
\begin{equation}
  \lambda_{\ell}^{TT}
  =
  \frac{\ell(\ell+D-1)-4}{L^2},
  \qquad
  \lambda_{\min}^{TT}
  =
  \frac{2(D-1)}{L^2}.
  \label{eq:ttgap}
\end{equation}
Thus, for normalized localized profiles, the observer becomes order-one on this benchmark TT channel when
\begin{equation}
  \frac{\gamma^2\norm{W}_{L^\infty,\mathrm{op}}^2}{2\mu^2}
  =
  O\!\left(\frac{D-1}{L^2}\right).
  \label{eq:scalar-orderone}
\end{equation}
This comparison should be read only as a geometric scale estimate. It does not derive the full conformal-sector dynamics and does not reproduce Maldacena's contour analysis. It shows that once the metric-coupled susceptibility becomes comparable to a low-lying gravitational scale, observer backreaction is no longer excluded by perturbative smallness.

\begin{figure*}[t]
\centering
\begin{tikzpicture}[x=1cm,y=1cm,every node/.style={font=\small}]

\node[font=\bfseries, align=center] at (-4.25,3.25)
  {(a) Projected scalar spectrum on $S^D$};
\node[font=\bfseries, align=center] at (4.25,3.25)
  {(b) Dimensionless susceptibility};

\begin{scope}[shift={(-7.6,0)}]
  \draw[->, thick] (0,0) -- (5.3,0) node[right] {$\ell$};
  \draw[->, thick] (0,0) -- (0,2.8) node[above] {$\lambda_\ell L^2$};
  \foreach \x/\lab in {0/0,1/1,2/2,3/3,4/4}
    \draw (\x,0.06) -- (\x,-0.06) node[below=3pt] {\lab};
  \draw[ultra thick, red!80!black] (-0.12,0.12) -- (0.12,-0.12);
  \draw[ultra thick, red!80!black] (-0.12,-0.12) -- (0.12,0.12);
  \node[anchor=north west, text=black, align=left] at (0.18,-0.08)
    {zero mode\\projected out};
  \foreach \x/\h in {1/0.75,2/1.35,3/1.95,4/2.45}{
    \draw[ultra thick, MidnightBlue] (\x,0) -- (\x,\h);
    \fill[MidnightBlue] (\x,\h) circle (1.6pt);
  }
  \draw[dashed, MidnightBlue] (0,0.75) -- (4.45,0.75);
  \node[anchor=west, text=black] at (4.52,0.75) {$\lambda_1=D/L^2$};
  \node[draw, rounded corners, fill=white, inner sep=3pt, align=left] at (2.55,2.35)
    {$\lambda_\ell=\ell(\ell+D-1)/L^2$\\[2pt]
     for $\ell\ge 1$ after projection};
\end{scope}

\begin{scope}[shift={(1.0,0)}]
  \draw[->, thick] (0,0) -- (6.0,0) node[right] {$x=m^2L^2/D$};
  \draw[->, thick] (0,0) -- (0,2.75);
  \node[rotate=90, align=center] at (-0.65,1.45)
    {$\displaystyle \frac{\|A_{L,m}^{-1}\|_{\perp 0}}{L^2/D}$};
  \foreach \x/\lab in {1/1,2/2,3/3,4/4,5/5}
    \draw (\x,0.06) -- (\x,-0.06) node[below=3pt] {\lab};
  \draw (0.06,2.2) -- (-0.06,2.2) node[left=4pt] {$1$};
  \draw[dashed, gray] (1,0) -- (1,2.2);
  \draw[domain=0:5.4, smooth, variable=\x, ultra thick, MidnightBlue]
    plot ({\x},{2.2/(1+\x)});
  \node[draw, rounded corners, fill=white, inner sep=3pt, align=center] at (1.55,2.35)
    {compactness-controlled\\$x\ll 1$};
  \node[draw, rounded corners, fill=white, inner sep=3pt, align=center] at (4.25,0.65)
    {mass-controlled\\$x\gg 1$};
  \node[draw, rounded corners, fill=white, inner sep=3pt, align=left] at (3.33,1.54)
    {$\displaystyle \|A_{L,m}^{-1}\|_{\perp0}=\frac{L^2}{D}\frac{1}{1+x}$};
\end{scope}

\end{tikzpicture}
\caption{
Quantitative visualization of the scalar infrared-susceptibility criterion.
In panel (a), the scalar zero mode on $S^D$ is removed, so the first nonzero eigenvalue is $D/L^2$ and the remaining spectrum stays discrete on a fixed compact sphere.
In panel (b), the exact inverse norm for $A_{L,m}=-\bar\nabla^2+m^2$ on the zero-mode-projected scalar sector is shown in dimensionless form,
$\|A_{L,m}^{-1}\|_{\perp0}=(L^2/D)(1+m^2L^2/D)^{-1}$.
The left regime is controlled by compactness, while the right regime is controlled by the mass scale.
Uniform semiclassical control fails only across families in which this susceptibility becomes parametrically large.
}
\label{fig:comparison}
\end{figure*}
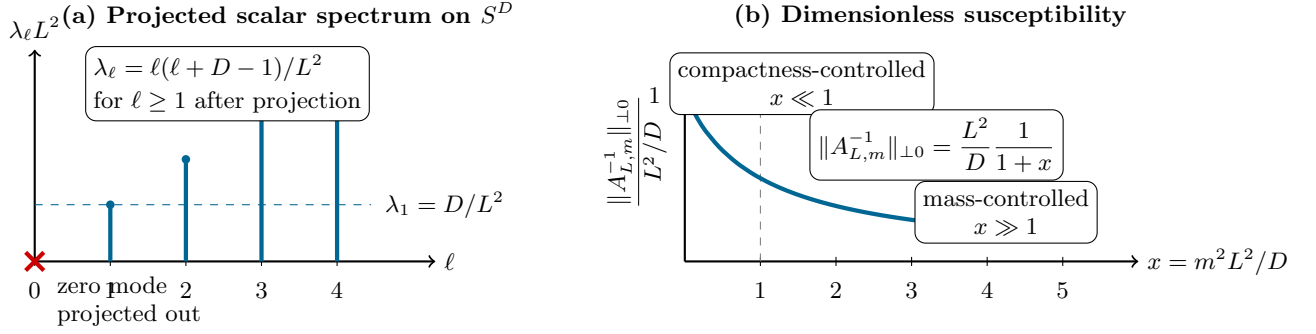

\paragraph{A self-contained observer taxonomy.---}
Independently of any microscopic observer interpretation, the Schur complement separates three cases.

First, an \emph{exact spectator} has vanishing mixed source on the channel under discussion:
\[
  \mathfrak j_P=0 .
\]
Its partition function may contain internal information, but it factorizes from the projected metric fluctuation at quadratic order.

Second, a \emph{perturbative spectator} has finite but small Schur correction:
\[
  \norm{A^{-1/2}\mathfrak j_P}^{2}
  \ll
  \delta_P,
\]
where \(\delta_P\) is the relevant stable gravitational scale on the projected channel. Such a sector renormalizes the Gaussian saddle but does not reorganize it.

Third, a \emph{gravitational observer relative to \(P\)} has finite but order-one metric-coupled susceptibility:
\[
  \norm{A^{-1/2}\mathfrak j_P}^{2}
  =
  O(\delta_P)
\]
on at least one relevant metric channel. This is the class capable of participating in a nontrivial saddle reorganization. It still does not automatically solve the de Sitter phase problem; in particular, if \(P\) is the conformal channel, contour and state-counting input remain essential.

The conceptual lesson is therefore sharper than a purely informational definition of observation. A worldline is not enough. An internal clock is not enough. A protected memory space is not enough. The observer must have controlled but non-negligible susceptibility to the metric directions that matter for the gravitational saddle.

\paragraph{\(\bm{\SU(3)}\) confinement as a conditional realization.---}
Among known physical mechanisms, confining \(\SU(3)\) dynamics are a natural candidate for finite infrared susceptibility. Confinement phenomenology and lattice calculations support a finite hadronic/glueball scale and short-ranged gauge-invariant correlators \cite{MorningstarPeardon1999}. In the strict mathematical sense, however, the Yang--Mills mass-gap problem on \(\mathbb R^4\) remains open and is one of the Clay Millennium Prize Problems \cite{ClayYangMills}. The present paper therefore uses \(\SU(3)\) only as a physically motivated conditional realization, not as a proved constructive derivation.

The conditional statement is precise. If a microscopic confining \(\SU(3)\) sector supplies, after gauge fixing, projection, and coupling to the relevant metric channel, a response operator with a positive infrared lower bound, then Proposition~1 applies. Sub-threshold fluctuations are spectators or perturbative spectators, while finite above-threshold excitations may backreact without generating an infrared catastrophe. In relation to the \(\SU(3)\) vacuum-atom framework \cite{Ali:2024rnw,Ali:2025wld}, the present paper identifies the semiclassical stability input that such a microscopic picture would have to provide: a bounded metric-coupled infrared susceptibility. It does not compute the full mixed block \(K\) for QCD, does not prove the Yang--Mills gap, and does not claim a nonperturbative construction of the observer kernel.

\paragraph{Broader physical context.---}
The distinction between exact spectators, perturbative spectators, and gravitational observers also clarifies the role of topological and memory-bearing systems. Topological field theories and topological orders can furnish stable logical sectors while remaining insensitive, after infrared reduction, to local metric fluctuations \cite{WenNiu1990QHTopDegeneracy,Kitaev2003ToricCode,QiHughesZhang2008TIFieldTheory}. Defect sectors in cosmology and condensed matter can store geometric information while remaining effectively decoupled from long-wavelength gravitational response \cite{Kibble1976TopologyStrings,VilenkinShellard1994,Zurek1985CosmoSuperfluidHelium}. These systems are informationally nontrivial, but they become gravitationally relevant only through the metric-coupled susceptibility measured by the Schur term.

\paragraph{Interpretation and conclusion.---}
The de Sitter phase problem can be restated in a sharper semiclassical language. The relevant Gaussian object is not merely the uncoupled gravitational Hessian, but the effective channel form obtained after the observer sector is integrated out. The induced correction is controlled by \(A^{-1/2}\mathfrak j_P\), or by \(K_P^\dagger A^{-1}K_P\) in the operator-bounded case. A spectral gap in the observer kernel gives bounded backreaction; metric-coupled soft modes destroy uniform control.

The present version goes beyond a purely necessary condition in one precise direction. If a projected channel is already stable, with \(Q_{gg}^{P}\ge\delta_P\norm{h}^{2}\), then the strict inequality \(\opnorm{A^{-1/2}\mathfrak j_P}^{2}<\delta_P\) is sufficient for the effective Gaussian channel to remain closed and coercive. The localized clock-detector model provides an explicit observer sector satisfying this condition whenever \(\Lambda_{\rm clk}^{2}/\Omega_0^2<\delta_P\). Thus the paper supplies both a necessary infrared-admissibility test and a concrete sufficient condition for stable-channel semiclassical control.

The limitations are equally important. The theorem does not solve the unrotated conformal-factor problem, does not reproduce Maldacena's contour or state-counting prescription, and does not prove that a confining \(\SU(3)\) sector supplies the complete microscopic observer kernel. In the conformal channel the estimates remain boundedness statements relative to an indefinite or externally contour-defined structure; no positivity conclusion follows without additional contour input. Maldacena-type phase cancellation uses this eligibility condition together with overlap with the relevant conformal or negative-mode sector and the independent contour or state-counting prescription; bounded susceptibility supplies the controlled observer input for that mechanism.

The novelty relative to Maldacena's construction is therefore not phase cancellation itself. It is the spectral filter that determines which observer sectors are eligible to enter a Maldacena-type cancellation within a controlled semiclassical expansion. Exact spectators carry information but do not couple to the relevant metric channel. Soft overlapping sectors couple, but destroy uniform control. Gapped sectors with bounded and non-negligible susceptibility define the admissible class on which contour and state-counting mechanisms can act.

This disciplined scope is also the source of the criterion's usefulness. It separates a purely informational observer from a gravitational observer by a physical and mathematical test. Observation is gravitationally relevant only when the observer sector couples to the metric with finite infrared susceptibility, and dynamically relevant only when that finite susceptibility is comparable to the gravitational scale of the channel. Stable macroscopic reality, in this semiclassical sense, requires neither arbitrary information storage nor uncontrolled soft response, but a threshold structure: enough coupling to participate in geometry, and enough spectral discipline to avoid catastrophic infrared backreaction.

\section*{Acknowledgments}
It is a pleasure to thank Bartek Czech, Roberto Emparan, Shoichiro Miyashita, Miguel Tierz, and Zhenbin Yang for illuminating discussions. RE gratefully acknowledges the hospitality of the Institute of Cosmos Sciences of the University of Barcelona (ICCUB) during the final stage of this work. RE is supported by the Dushi Zhuanxiang Fellowship and acknowledges a Shuimu Scholarship as part of the Shuimu Tsinghua Scholar Program.

\bibliographystyle{apsrev4-2}
\bibliography{refs}

\end{document}